\documentclass[letter]{aa} % for a referee version
\usepackage{graphicx}
\begin{document}

   \title{Orbital features of distant trans-Neptunian objects induced by giant gaseous clumps}

   \author{V.V.Emel'yanenko
          \inst{}}

   \institute{Institute of Astronomy, Russian Academy of Sciences,
              Moscow, 119017 Russia\\
              \email{vvemel@inasan.ru} }

   \date{Received                       ; accepted                          }

% \abstract{}{}{}{}{} 
% 5 {} token are mandatory
 
  \abstract
   {The discovery of distant trans-Neptunian objects  has led to heated discussions about the structure of the outer Solar System.}
   {We study the dynamical evolution of small bodies from the Hill regions of migrating giant gaseous clumps that form in the outer solar nebula via gravitational fragmentation. We attempt to determine whether the observed features of the orbital distribution of distant trans-Neptunian objects could be caused by this process.}
   { We consider a simple model that includes the Sun, two point-like giant clumps  with masses of $\sim 10$ Jupiter masses, and a set of massless objects initially located in the Hill regions of these clumps. We carry out numerical simulations of the motions of small bodies under gravitational perturbations from two giant clumps that move in elliptical orbits and  approach each other. The orbital distribution of these small bodies is compared with the observed distribution of distant trans-Neptunian objects.}
   {In addition to the known grouping in longitudes of perihelion, we note new features for observed distant trans-Neptunian objects. The observed orbital distribution points to the existence of two groups of distant trans-Neptunian objects with different dynamical characteristics. We show that the main features of the orbital distribution of distant trans-Neptunian objects can be explained by their origin in the Hill regions of migrating giant gaseous clumps. Small bodies are ejected from the Hill regions when the giant clumps move in high-eccentricity orbits and have a close encounter with each other.}
   {The resulting orbital distribution of small bodies in our model and the observed distribution of distant trans-Neptunian objects  have similar features.}

   \keywords{Kuiper belt: general --
                planet-disk interactions --
                methods: numerical
               }
\titlerunning{Orbital features of distant trans-Neptunian objects}
  \maketitle

%
%-------------------------------------------------------------------

\section{Introduction}

The discovery of distant trans-Neptunian objects (TNOs) moving in orbits with semimajor axes $a >150$ au  and perihelion distances $q>30$ au has provided unexpected information about the structure of the outer Solar System. \citet{TS14} first suggested that there is a concentration of arguments of perihelion $\omega$ near the value $\omega = 0^\circ$. This value corresponds to one of the centers of the libration zones in the von Zeipel-Lidov-Kozai mechanism of secular perturbations  \citep{Ze910,Li62,Ko62}. Therefore, an assumption was made about the existence of a distant planet producing this effect. Later, \citet{BB16} noted that, to a greater extent, there is a grouping of longitudes of ascending nodes $\Omega$ and  longitudes of perihelion $\pi =\omega+\Omega$ , and this effect is associated with the combined action of orbital and secular resonances \citep{BM17}. Although \citet{Laetal17}  and  \citet{Shetal17b} scrutinized observational bias in the Outer Solar System Origins Survey sample and
concluded that there is no evidence of clustering in angular orbital element distributions  \citep[see also][]{Beetal20,CK20,Tr20}, 
\citet{Br17} estimated a rather low probability for the observed
sample of TNOs with $a > 230$ au to be drawn from a uniform
population. Detailed studies have shown that the observed features in the distribution of angular elements for distant TNOs could be produced by a planet with a mass of   $\sim 5\--10$ Earth masses moving in an orbit with a semimajor axis  $a \sim 400\--800$ au, eccentricity $e \sim 0.2\--0.5$, and inclination $i \sim 15\--25^{\circ}$  \citep{BB16,Baetal19}.

Although the dynamical arguments seem quite convincing
\citep{BM17}, the question
of the actual existence of a ninth planet in the Solar
System remains open. Despite ongoing searches, the
planet has not yet been discovered. An explanation for
the formation of such a massive and distant planet is
also difficult to establish \citep{Baetal19}.  \citet{Shetal17a} and \citet{Kaetal20} have found dynamical
effects that a massive distant planet would have
on the distant TNO population that were not
highlighted in the initial published theory of \citet{BB16}. 
\citet{Neetal17} concluded that the inclination distribution of
Jupiter-family comets is wider than the observed one in models with a ninth planet.

However, without existing additional perturbers, it is difficult to explain the observed clustering of the apsidal lines of distant TNOs. It has been suggested that a self-gravitating massive disk of TNOs, and not a ninth planet, may be sustaining the confinement of the longitudes of perihelion \citep{ST19,Zdetal20}. This suggestion was criticized in \citet{Baetal19}. Their main objections concern the lack of material needed for the formation and existence of the massive disk of TNOs. The arguments discussed were based on the widespread "core accretion" theory of planet formation \citep[e.g.,][]{Sa91,Moetal06,Ar18}  and the classical "minimum mass solar nebula" model of the solar protoplanetary disk \citep[e.g.,][]{We77,Ha81,De07,Cr09}.

An alternative planet formation scenario is provided by the "gravitational instability" theory \citep[see, for example, the reviews][]{KL16,Na17}. The detection of exoplanets in wide orbits provides
observational support for this theoretical scenario \citep[e.g.,][]{Maetal08,Boetal20}. 
The gravitational instability of protoplanetary disks is envisioned to lead to
the formation of self-gravitating clouds of gas and dust (the canonical gas-to-dust mass ratio is 100:1), called giant gaseous clumps.
 Recently, it has been
shown that such clumps  participate in a
complex dynamical interaction with the disk, which notably
leads to the migration of clumps \citep{Maetal02, VB05,Na10,Baetal11,Zhetal12,St15,VE18}. Moreover,
migrating clumps can experience close encounters
with each other, which often lead to the ejection of
objects into hyperbolic orbits \citep{TP02,VE18}. 
\citet{Em20} showed that the observed  distribution of angular orbital elements of TNOs could be created by the perturbations of such clumps, based on the results of hydrodynamical simulations  in \citet{VE18}.

 In this paper, we first  discuss new features of the observed distant TNOs. Then we consider dynamical processes of migration and close encounters of giant gaseous clumps, motivated by the results of \citet{VE18}. We attempt to determine whether the observed orbital distribution of distant TNOs could be induced by  such clumps moving in orbits with different inclinations to the ecliptic. For this purpose, we study the dynamical evolution of small bodies from the Hill regions of the giant clumps.  In the numerical hydrodynamics simulations of  \citet{VE18}, the first clump forms 60 kyr after the star formation, and no clumps are present in the protoplanetary disk at 0.5 Myr. Thus, our model relates to a very early stage in the evolution of the protoplanetary disk when the present-day planets were not yet formed.

\section{Orbital features of the observed distant trans-Neptunian objects}         

We consider TNOs with $q > 40$ au, in contrast to many previous works
\citep[e.g.,][]{TS14,BB16} that analyzed orbits with $q > 30$ au.
This restriction seems more justified when
studying dynamical mechanisms in the early stages of Solar System
formation. Orbits with perihelia near the orbit of Neptune have changed significantly over the lifetime of the Solar System due to planetary  perturbations. Moreover, the stronger
restriction on perihelion distances allows us to take into consideration not only orbits with very large $a$, 
but also a population of detached TNOs with  smaller $a$.  The heliocentric orbital elements presented
on the website of the Minor Planet Center\footnote[1] {https://www.minorplanetcenter.net/iau/lists/} on
January 16, 2020, were used for objects
observed in at least two oppositions.  A list of these objects and their orbital elements are given in Appendix A.

Figure~\ref{fig1} shows the distribution of semimajor axes and inclinations for TNOs 
with $q>40$ au and $a>60$ au. All the objects with $a<160$ au have $i>20^\circ$. Large inclinations are not
very unusual because there are dynamical mechanisms, in particular resonances, by which TNOs
can reach high-inclination orbits. Really surprising is the lack of low-inclination orbits for all objects with $q>40$ au and $a>60$ au. Objects with 
larger semimajor axes have lower inclinations on average, and all the objects with $a>400$ au have 
$i<15^\circ$. 
Figure~\ref{fig2} shows the distribution of longitudes of perihelion and perihelion distances 
for TNOs with $q>40$ au and $a>100$ au.  For $q>40$ au, two concentrations 
of longitudes of perihelion are clearly visible, even for objects with $a>100$ au. The first group of 12 objects 
has the mean value $\overline{\pi}=66^\circ$, and the second group of six objects has the mean value 
$\overline{\pi}=266^\circ$. The observed values of perihelion distances for these groups are different.
For the first group, perihelion distances extend from 44 au to 80 au. Perihelion distances for the second group are 
smaller; they do not exceed  45 au. The orbits in these groups also differ in regards to the distribution of inclinations.
While the inclinations for the first group are in the interval $(4^\circ, 20^\circ)$ with the mean value 
$\overline{i}=15^\circ$, the inclinations for the second group spread from $14^\circ$ to $48^\circ$ with
the mean value $\overline{i}=30^\circ$. The described features give us reason to suppose that we are observing two groups of objects with different dynamical characteristics. In this paper, we assume that these features are inherent in the real distribution of distant TNOs and are not  generated by observational biases.

   \begin{figure}
   \centering
      \resizebox{\hsize}{!}{\includegraphics{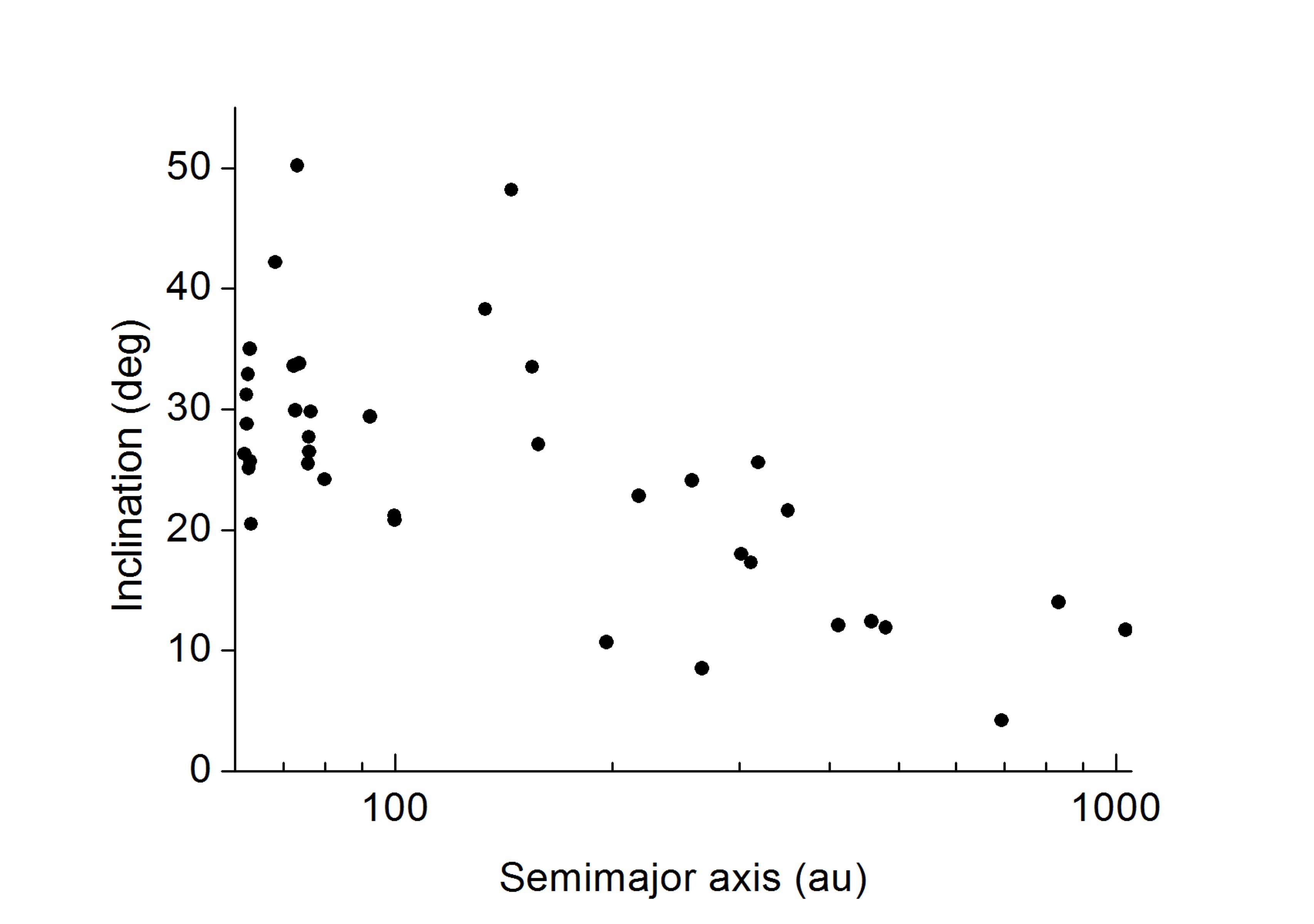}}
      \caption{Distribution of semimajor axes and inclinations for observed 
      multiple-opposition TNOs with $q>40$ au and $a>60$ au.}
         \label{fig1}
   \end{figure}

 \begin{figure}
   \centering
      \resizebox{\hsize}{!}{\includegraphics{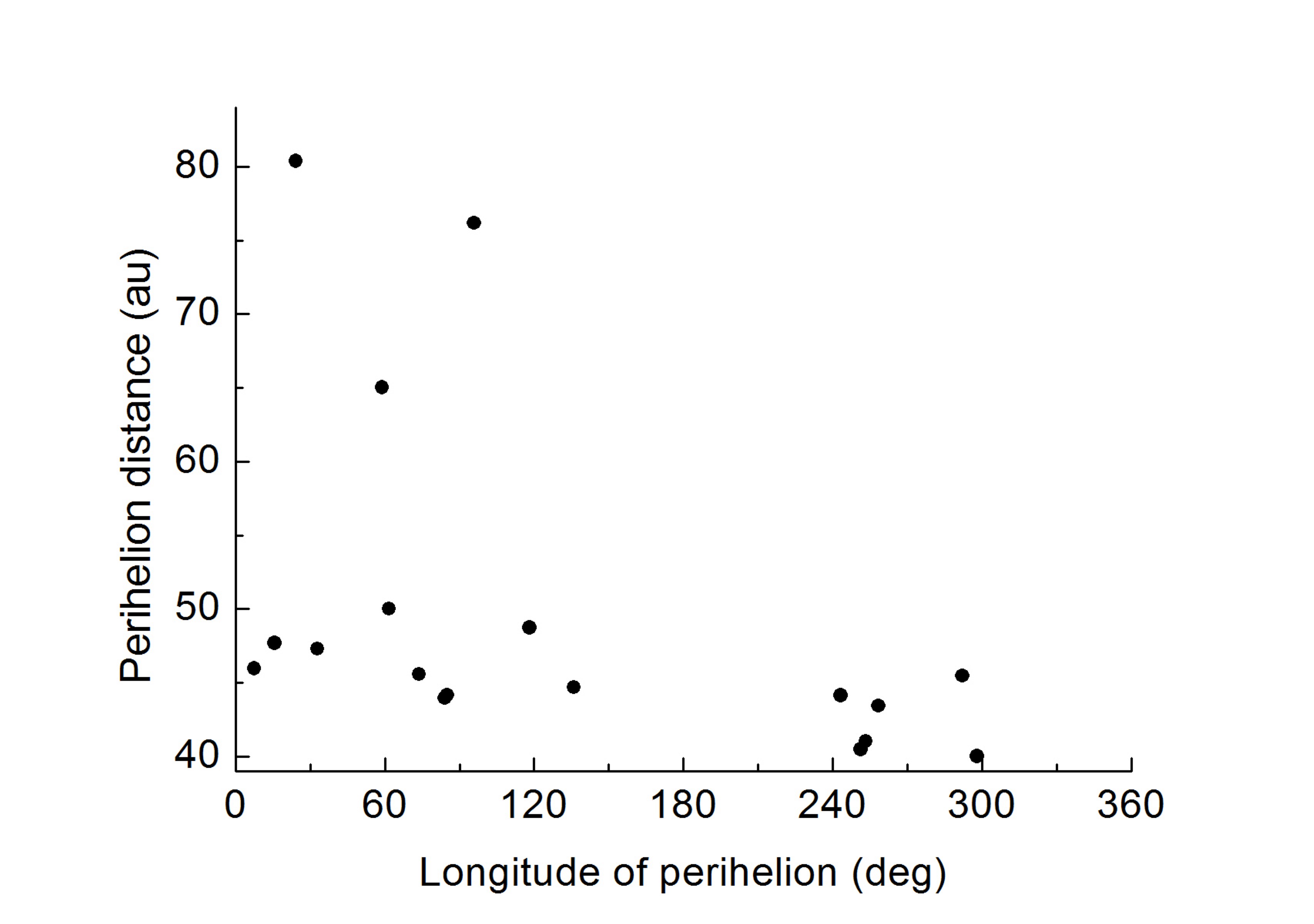}}
      \caption{Distribution of longitudes of perihelion and perihelion distances for observed 
      multiple-opposition TNOs with $q>40$ au and $a>100$ au.
              }
         \label{fig2}
   \end{figure}
   
\section{Methods}
 
We study a system of two interacting gaseous clumps surrounded by planetesimals in 
the Hill regions  defined by the radius 
$R_\mathrm{H} \approx r_h\left((m/3)^{1/3}-\frac{1}{3}(m/3)^{2/3}-\frac{1}{9}(m/3)\right)$, where $r_h$ is the clump heliocentric distance and $m$ is the clump mass in the astronomical system of units 
\citep[e.g.,][]{MuDe99}. This approach is based on the results from \citet{VE18} regarding the dynamics of clumps. In this model, the outer object with
a lower mass moves in an initial orbit with  $e \sim 0.5$.
The inner object starts moving in a near-circular
orbit with $a \sim 100$ au. After a certain period of time, the objects approach each other.
As a result of
mutual perturbations, these objects enter into orbits with
high eccentricities. The outer object is ejected to a  near-parabolic or
hyperbolic orbit. The inner object enters
an orbit with $q \sim 30$ au, and this orbit is quickly
rounded off at such a distance due to interaction with
the protoplanetary gas disk \citep[see Fig. 17 in][]{VE18}. 

We assume that planetesimals exist in the Hill regions of
giant clumps during their migration stage. According to the calculations in \citet{VE18}, the lifetimes of giant
clumps in the protoplanetary disk do not exceed
0.5 Myr. Although the question of the formation of
small bodies in such a short time period is open, there
are many works that consider a very rapid formation
of small bodies in the outer Solar System. For example,
\citet{WJ14} claim that planetesimals of various sizes can be made at
40 au in their numerical experiments through the collapse of pebble clouds formed by 
the streaming instability  in protoplanetary disks: more than $\sim100$ km
in $\sim 25$ years, 10 km in several hundred years, and 1 km in
several thousand years. \citet{Boetal10}, \citet{NC12}, and \citet{Na17} argue that making solid bodies is more straightforward 
inside the giant gaseous clump regions than in circumstellar disks.

The main dynamical mechanism of creating the population  of small bodies
moving in distant heliocentric orbits is associated with a decrease in the size of 
the Hill region when the clump moves from aphelion to perihelion. Some bodies formed in the Hill region  
 are outside this region at a certain time and are ejected from the vicinity of the clump.
 
Here we consider simple models in which the clump is treated
as a point mass and the planetesimals are treated as massless particles.
The particles are initially placed in the plane of the heliocentric motion of the clump.  
The planetesimal disk has the initial surface density profile of 
$\Sigma(r) \propto 1/\sqrt{1+(r/r_0)^4}$   \citep{VE19},
where $r$ is the radius of a circular orbit around the center of the clump. 
According to the numerical simulations in \citet{VE19}, $r_0$ is usually much less than $R$, where $R$ is the clump radius; then, $\Sigma(r)$ is approximately
proportional to $r^{-2}$ for $r>R$.  For each $r$, the initial particles  have a random uniform distribution 
along the orbit. We consider particles with initial  $r>R$ and prograde 
Keplerian velocities.

The integration of the equations of motion is based on the symplectic integrator 
for the $N$-body problem \citep{Em07}. To describe
the changes of orbits during the migration of clumps, we applied a method used
in \citet{Em11}. 

\section{Numerical simulations}

We consider a simple model that includes the Sun, two point-like massive clumps, and a set of particles initially located  in the Hill regions of these clumps.
To model the motion of clumps, we relied on the
results of  \citet{VE18}, but we carried out numerical simulations with different masses and orbits 
of clumps. Below we demonstrate variants for which the basic features of 
the  simulated distribution are similar to those observed for distant TNOs. 
All the analyzed orbital elements are heliocentric. 

In the first stage of our model, two clumps migrate slightly inward. The outer clump, 
with a  mass of  12 Jupiter masses and radius $R=6.4$ au, 
starts in the aphelion of the orbit with $a=187$ au, $q=123$ au, $i=30^\circ$, and $\pi=257^\circ$. 
The inner object,  with a mass of  17 Jupiter masses and radius $R=7.2$ au, starts 
in the orbit with $a=110$ au, $q=100$ au, $i=15^\circ$, and $\pi=98^\circ$.  Near every clump, 1000 particles are initially distributed between $R$ and $R_\mathrm{H}$ at the  aphelion distances of the clumps (37 au for the outer clump and 20 au for the inner clump). The dynamical evolution of the clumps is shown in Fig.~\ref{doi} in Appendix B. The orbital distribution of particles in the first stage (after 2300 years) is described in Appendix C.

In our model, there is then a close encounter of the clumps. This encounter occurs near 
the perihelion of the outer clump and the aphelion of the inner clump. As a result, the inner 
clump evolves to have an orbit with $q=37$ au. According to \citet{VE18}, the orbit quickly becomes near-circular 
at such a distance due to interaction with
the protoplanetary gas disk, and the clump loses its mass. After that, this object weakly disturbs 
distant particles. Therefore, we consider the orbital distribution of particles near the moment
of the perihelion passage of the inner clump.

\begin{figure}[h]
   \centering
   \resizebox{\hsize}{!}{\includegraphics{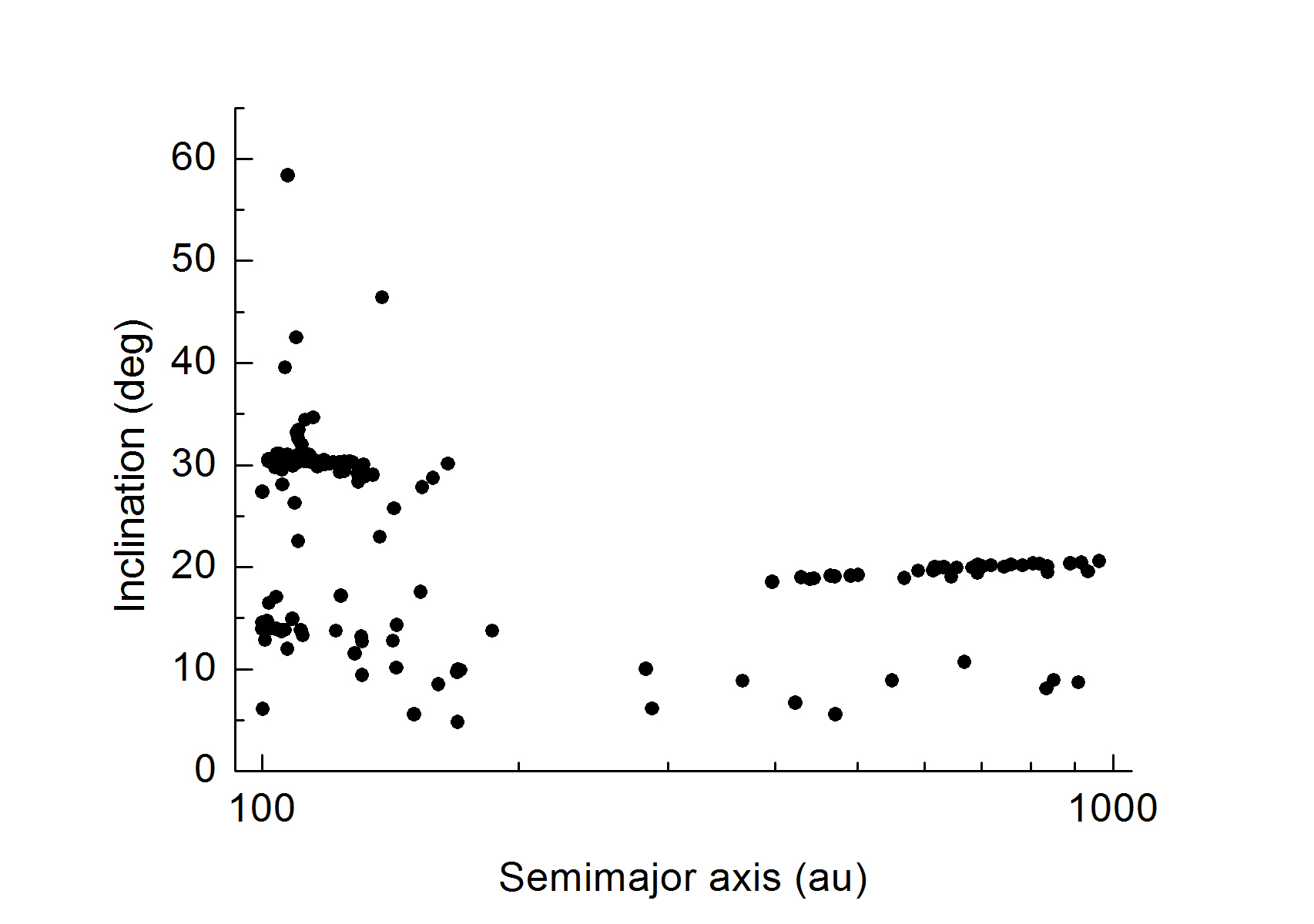}}
      \caption{Distribution of semimajor axes and inclinations for simulated particles with 
$40<q<80$ au and $100<a<1000$ au.}
         \label{fig3}
   \end{figure}

\begin{figure}[h]
   \centering
      \resizebox{\hsize}{!}{\includegraphics{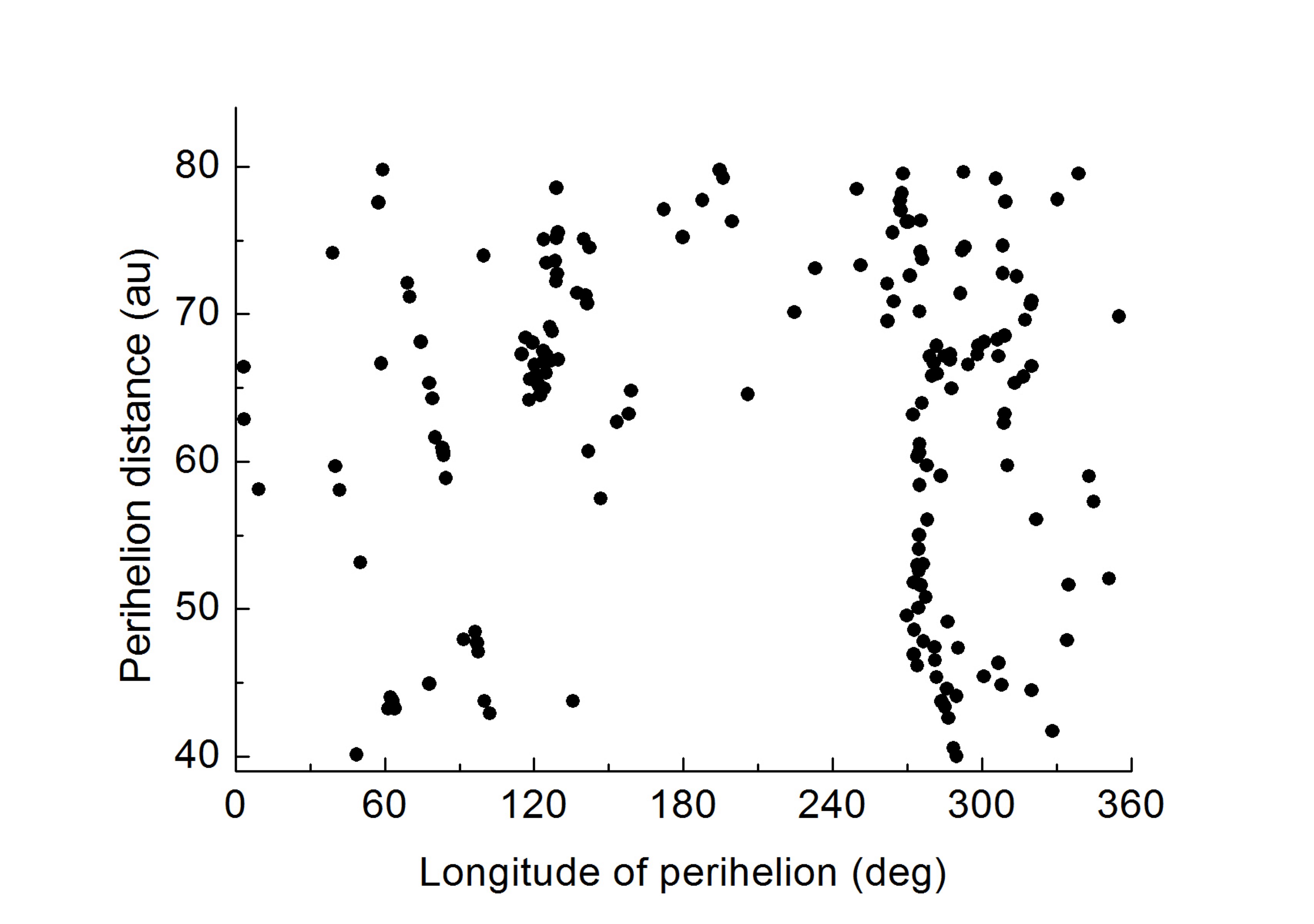}}
      \caption{Distribution of longitudes of perihelion and perihelion distances for simulated particles with 
$40<q<80$ au and $100<a<1000$ au.}
         \label{fig4}
   \end{figure}

Figure \ref{fig3} shows the distribution of semimajor axes and inclinations
for particles with $40<q<80$ au and $100<a<1000$ au (the observable region of distant TNOs). 
Only particles located outside the Hill regions
of the clumps are included. Figure~\ref{fig4} shows the distribution of longitudes of perihelion and perihelion distances 
for the same particles. The main features of these distributions are consistent with those shown in Figs. 1 and 2.
On average, the inclinations of the orbits decrease as the semimajor axes increase. There are two concentrations of 
particles in the distribution of longitudes of perihelion (applying the Kolmogorov-Smirnov statistical test to those   presented in Fig.~\ref{fig4} shows that the hypothesis of a uniform distribution is rejected with a probability of more than 0.9999). We note that orbits of these two groups differ in inclinations (for example, the mean inclination of orbits with $\pi<180^\circ$ is equal to $16.4^\circ$, while the mean inclination of orbits with $\pi>180^\circ$ is equal to $28.7^\circ$). 

\section{Discussion}

We have studied a model in which two giant gaseous clumps approach each other and move into high-eccentricity orbits.
In this paper, we do not consider the possible further evolution of such clumps. According to  \citet{VE18}, the inner clump is rapidly destroyed due to tidal effects. In the model discussed above, the final orbit of the outer object has $q=106$ au and $a=544$ au. It is natural to assume that this clump also gradually disperses in such an orbit, adding to the swarm of solid bodies in the distant Solar System. In some variants of our simulations, such a clump is ejected into a hyperbolic orbit, as in the work of \citet{VE18}.
        
\begin{figure}[h]
   \centering
      \resizebox{\hsize}{!}{\includegraphics{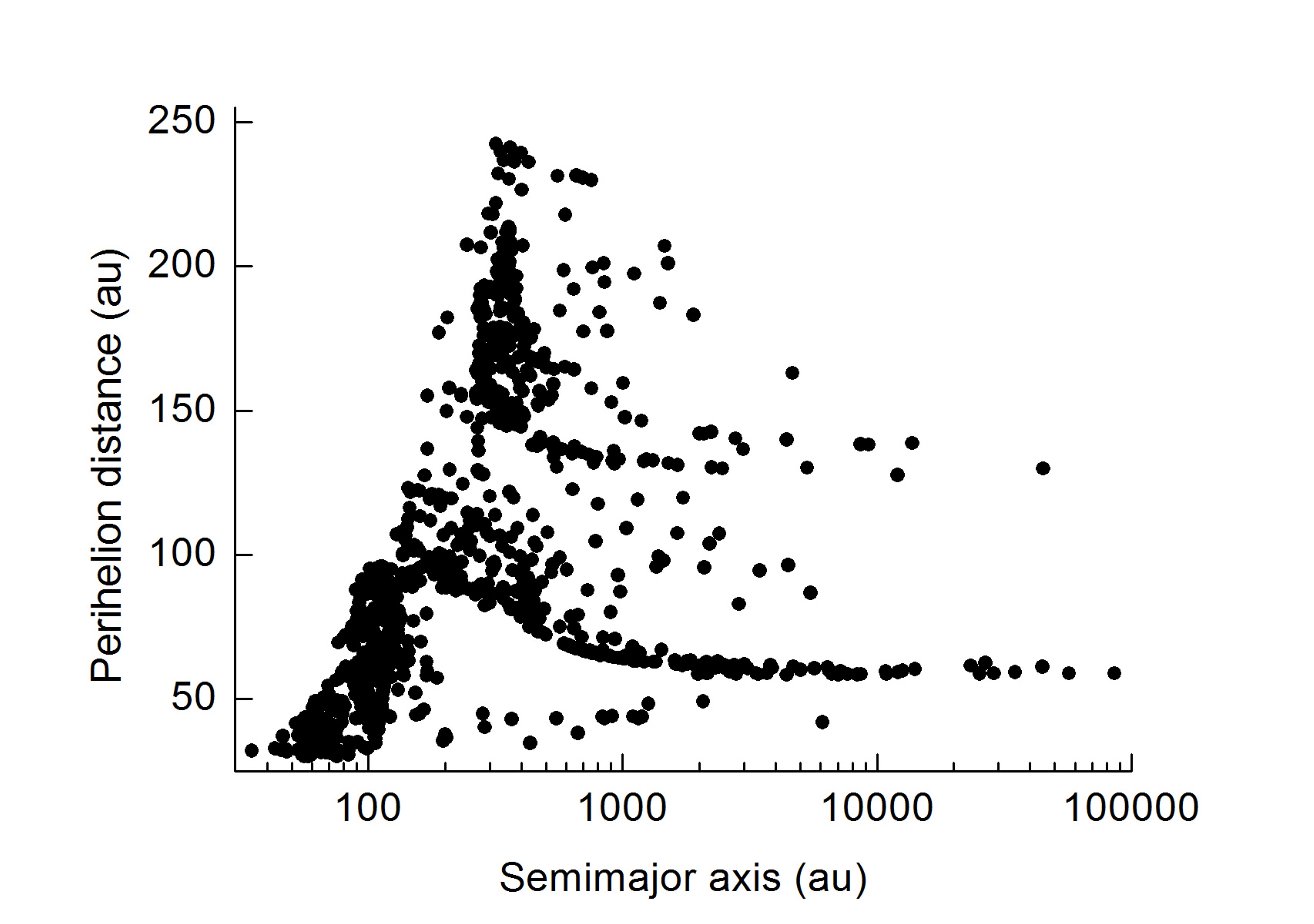}}
      \caption{Distribution of semimajor axes and perihelion distances for simulated particles with 
$q>30$ au and positive $a<10^5$ au.}
         \label{fig5}
   \end{figure}

The clustering of orbital elements  is clearly discernible in our simulations. But this clustering should be sustained by some mechanism during the lifetime of the Solar System in order to be seen in the present observational data. For example, planetary perturbations are large enough to disperse the initial clustering of longitudes of perihelion throughout the lifetime of the Solar System, even for Sedna-type objects \citep[e.g.,][]{Saetal19}. \citet{ST19} propose that a self-gravitating disk of TNOs can maintain its orbital features throughout the lifetime of the Solar System. This scenario is very relevant for our model. Figure~\ref{fig5} shows the distribution of semimajor axes and perihelion distances for all particles with $q>30$ au and  positive $a<10^5$ au. Obviously, it is not difficult to obtain  the required mass of  the disk of $\sim 10$ Earth masses \citep{Baetal19,ST19} if the initial giant clumps have masses of  $\sim 10$ Jupiter masses. We leave a detailed study of the orbital evolution of the disk population over the lifetime of the Solar System to a future paper.
        
It is difficult to discuss in more detail the consistency of the suggested dynamical process and the observed distribution of distant TNOs. On the one hand, the number of detected distant TNOs is currently too small. On the other hand, it is evident that the process of the formation and migration of giant gaseous clumps is much more complicated than in our model. In particular, while we consider only two clumps on a relatively short timescale, numerical hydrodynamics simulations \citep[e.g.,][]{Vo13,VP17,VE18} show that the formation process of giant gaseous clumps continues for $\sim 0.5$ Myr, with many clumps involved. It is actually surprising that the studied simple model describes the main features of the observed orbital distribution of distant TNOs so well. 

In addition to the abovementioned features, Fig.~\ref{fig5} shows that there is a gap in the distribution of perihelion distances in the observable region for $a>170$ au. A similar gap exists for the observed distant TNOs (although there are also no observed TNOs with  $q>56$ au for smaller semimajor axes). This figure also shows that our model produces many objects with $30<q<40$ au and high eccentricities. Although modern theories relate the origin of such objects to the result of gravitational scattering by Neptune (the so-called scattered disk) \citep[e.g.,][]{ MN20}, our numerical simulations  demonstrate that giant clumps migrating in the outer Solar System could play a significant role in creating this population of TNOs.
 
Future observations confirming the proposed dynamical process would have important implications for understanding the structure of the outer Solar System. We would like to note three points that were not explicitly discussed in the paper but follow from the general approach of the considered model.

First, the solar protoplanetary disk was more extended than the classical minimum mass
solar nebula model formulates. The theory of a massive, extended solar nebula has long existed, and it is supported by observations of massive disks around pre-main sequence and main sequence stars \citep[e.g.,][]{ Ba94}.

Second, there are many undiscovered small bodies with $q>80$ au (see Fig.~\ref{fig5}). A significant additional contribution can come from the disintegration of the outer clump.

 Thirdly, Fig.~\ref{fig5} shows that many objects in our model reach high-eccentricity orbits located in both the inner core ($10^3<a<10^4$ au)  \citep{Hi81} and the outer component ($a>10^4$ au) of the Oort cloud \citep{Oo50}. Therefore, the formation of the Oort cloud may be associated
with the described process, although the main models relate the origin of  the Oort cloud to the scattering of remaining planetesimals by the formed planets \citep[e.g.,][]{Fe80,Duetal87,Doetal04,Emetal07}.

\section{Conclusions} 

We have analyzed the orbital distribution of distant TNOs. We note several new features of the observed objects in addition to their known groupings in longitudes of perihelion. Although the number of detected objects is still too small to make final conclusions, the observed orbital distribution points to the existence of two groups of distant TNOs with different dynamical characteristics.

We have shown that the main features of the observed orbital distribution of distant TNOs can be explained by their origin in the Hill regions of migrating giant gaseous clumps that form in the outer solar nebula via gravitational fragmentation. In our model, two clumps with masses of $\sim 10$ Jupiter masses interact gravitationally with the small bodies that surround them. The small bodies are ejected from the Hill regions when two giant clumps move in high-eccentricity orbits and have a close encounter with each other. The resulting orbital distribution of small bodies in our model and the observed distribution of distant TNOs have similar features.

\begin{acknowledgements}
 The author acknowledges the support of the Large Scientific Project of the Russian Ministry of Science and Higher Education "Theoretical and experimental studies of the formation and evolution of extrasolar planetary systems and characteristics of exoplanets" (project No.  13.1902.21.0039). The calculations were carried out using the MVS-100K supercomputer of the Joint Supercomputer Center of the Russian Academy of Sciences.  The author thanks an
anonymous referee for helpful comments.
\end{acknowledgements}


\begin{appendix}

\section{Orbital elements of distant multiple-opposition TNOs}
Table~\ref{table1} gives the heliocentric orbital elements and their standard deviations for the TNOs with $a>100$ au used 
in Sect. 2. These values were taken from the Jet Propulsion Laboratory website\footnote[2]{https://ssd.jpl.nasa.gov/sbdb.cgi} on September 2, 2020.

\begin{table*}[h]
\caption{Heliocentric orbital elements of distant multiple-opposition TNOs at epoch 2020 May\ 31.0.} % title of Table
\label{table1} % is used to refer this table in the text
\centering % used for centering table
\begin{tabular}{c c c c c c} % centered columns (6 columns)
\hline\hline % inserts double horizontal lines
Object & $a$ & $q$ & $i$ & $\omega$ & $\Omega$  \\ % table heading
           &  (au) & (au) & (deg) & (deg) & (deg) \\
\hline % inserts single horizontal line

2000 CR105 &216.1$\pm$0.6&44.076$\pm$0.005&22.8207$\pm$0.0006&316.38$\pm$0.01&128.3726$\pm$0.0003 \\
2003 VB12 &484.4$\pm$0.3&76.257$\pm$0.005&11.9307$\pm$0.0000&311.35$\pm$0.01&144.2479$\pm$0.0011 \\
2004 VN112 &326.7$\pm$1.9&47.288$\pm$0.007&25.5851$\pm$0.0003&326.73$\pm$0.01&65.9868$\pm$0.0005\\
2010 GB174 &342.3$\pm$22.5&48.645$\pm$0.310&21.5983$\pm$0.0052&347.13$\pm$0.36&130.8929$\pm$0.0198 \\
2012 VP113 &261.5$\pm$1.4&80.399$\pm$0.089&24.1082$\pm$0.0023&293.57$\pm$0.37&90.6829$\pm$0.0056 \\
2013 FT28 &304.8$\pm$10.1&43.394$\pm$0.113&17.3579$\pm$0.0034&40.78$\pm$0.16&217.7497$\pm$0.0048 \\
2013 GP136 &153.9$\pm$0.2&41.009$\pm$0.009&33.5230$\pm$0.0006&42.70$\pm$0.04&210.7320$\pm$0.0001 \\
2013 RA109 &479.5$\pm$2.3&45.977$\pm$0.015&12.4053$\pm$0.0001&262.79$\pm$0.15&104.6721$\pm$0.0055 \\
2013 SY99 &728.0$\pm$25.0&50.078$\pm$0.056&4.2223$\pm$0.0012&31.83$\pm$0.11&29.5104$\pm$0.0052 \\
2013 UT15 &198.9$\pm$0.8&44.049$\pm$0.026&10.6508$\pm$0.0010&251.87$\pm$0.03&191.9448$\pm$0.0004 \\
2014 SR349 &311.9$\pm$19.9&47.680$\pm$0.283&17.9686$\pm$0.0017&340.76$\pm$0.63&34.8624$\pm$0.0149\\
2014 SS349 &149.5$\pm$1.2&45.290$\pm$0.060&48.1325$\pm$0.0025&147.62$\pm$0.14&144.1310$\pm$0.0011\\
2015 KE172 &132.7$\pm$0.1&44.141$\pm$0.002&38.3204$\pm$0.0010&15.75$\pm$0.08&227.5674$\pm$0.0003\\
2015 KG163 &805.0$\pm$5.9&40.493$\pm$0.008&13.9822$\pm$0.0012&32.29$\pm$0.10&219.0906$\pm$0.0017\\
2015 KH163 &157.9$\pm$0.6&39.956$\pm$0.026&27.0854$\pm$0.0014&230.58$\pm$0.05&67.5927$\pm$0.0006\\
2015 RX245 &429.1$\pm$36.3&45.635$\pm$0.230&12.1244$\pm$0.0028&64.91$\pm$0.30&8.5925$\pm$0.0002\\
2015 TG387 &1085.5$\pm$111.5&65.160$\pm$0.212&11.6545$\pm$0.0006&117.78$\pm$0.32&300.7795$\pm$0.0072\\
2018 VM35 &251.9$\pm$64.0&44.962$\pm$1.282&8.4808$\pm$0.0034&302.91$\pm$2.68&192.4070$\pm$0.0584 \\
\hline %inserts single line
\end{tabular}
\end{table*}
\section{Dynamical evolution of the studied clumps}
\begin{figure}[h]
   \centering
   \resizebox{\hsize}{!}{\includegraphics{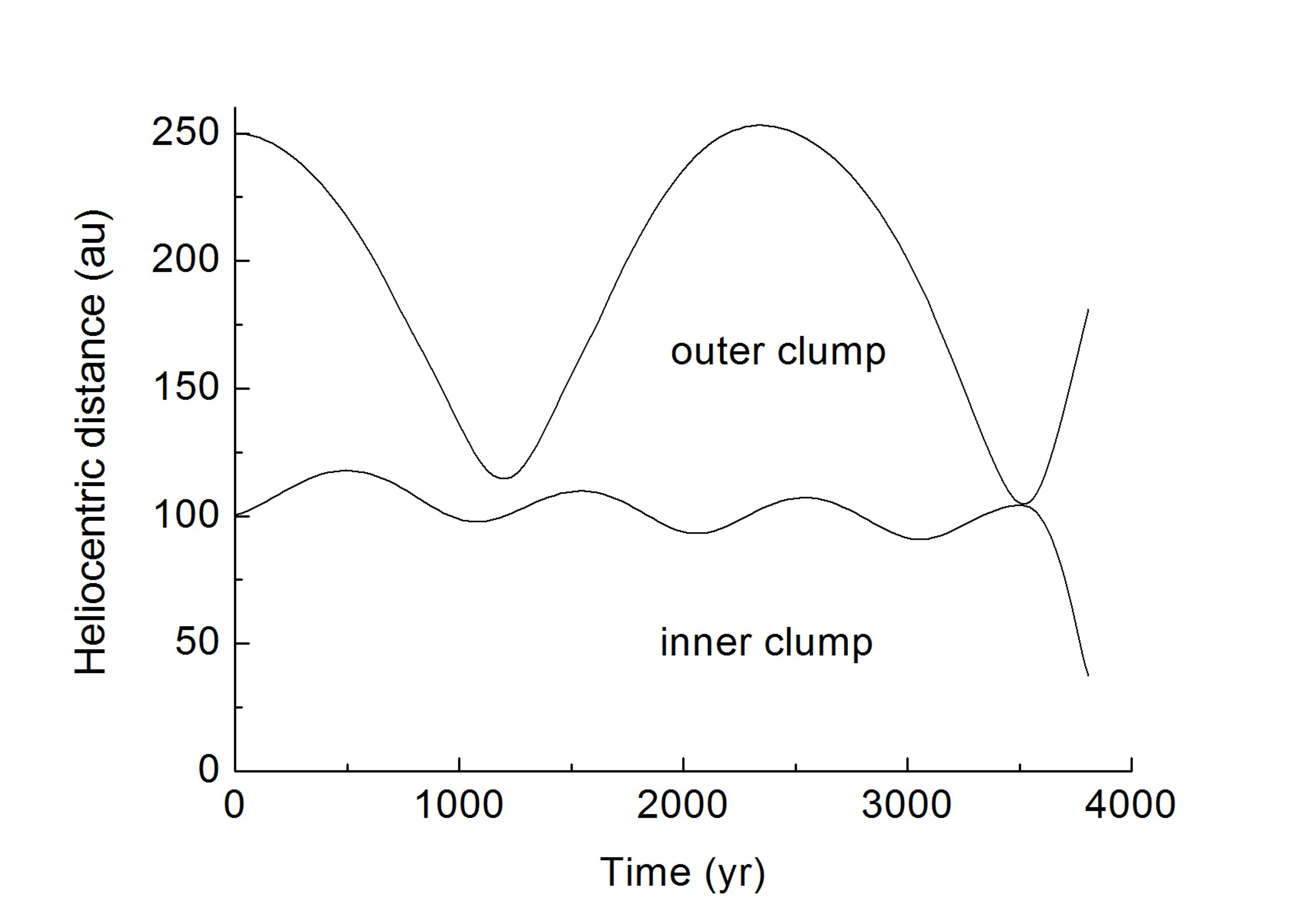}}
      \caption{Heliocentric distances of two clumps.}
         \label{doi}
   \end{figure}

Figure~\ref{doi} shows 
the heliocentric distances  of  two clumps as a function of time.
 
\section{Orbital distribution of particles before a close encounter}

\begin{figure}[h]
   \centering
      \resizebox{\hsize}{!}{\includegraphics{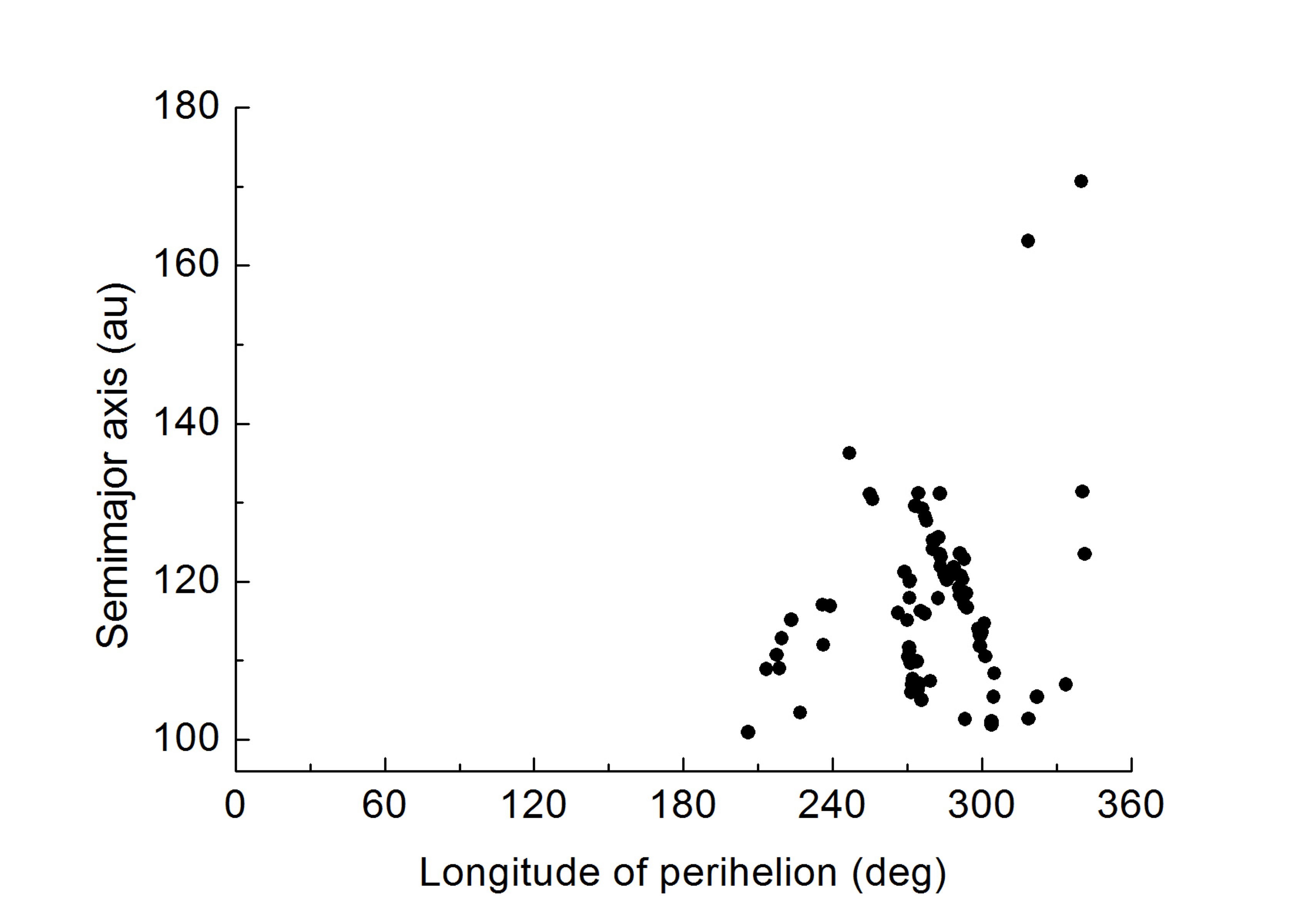}}
      \caption{Distribution of longitudes of perihelion and semimajor axes for simulated particles 
 with $40<q<80$ au and $a>100$ au after the first stage.}
         \label{figc}
   \end{figure}

After 2300 years of evolution, the outer clump reaches the orbit with $a = 180$
au and $q = 107$ au, and the inner clump moves in the orbit with $a = 100$ au and the aphelion
distance $Q = 107$ au. Figure~\ref{figc} shows the distribution of longitudes of perihelion and semimajor axes for particles  
with $40<q<80$ au and $a>100$ au (the observable region of distant TNOs) at this time. 
Only particles located outside the Hill region are included. 
All these particles originate in the Hill region of the outer clump. The inner clump
does not contribute to the observable region at this stage. The longitudes of perihelion of the particles 
are concentrated near the value of $\pi$ of the outer clump.
 
\end{appendix}

\end{document}